\documentclass[twocolumn,showpacs,amsmath,amssymb,prl]{revtex4}

\usepackage{graphicx}

\begin{document}

\title{Stationary state volume fluctuations in a granular medium}
\author{Matthias Schr\"oter}
\email{schroeter@chaos.utexas.edu}
\author{Daniel I. Goldman} 
\author{Harry L. Swinney} 
\affiliation{Center for Nonlinear Dynamics and Department of Physics\\ University of Texas at Austin, Texas 78712, USA}

\date{\today}

\begin{abstract}
A statistical description of static granular material requires  ergodic sampling of  
the phase space spanned by the different configurations of the particles. 
We periodically fluidize a column of glass beads and  find that  the sequence of volume fractions $\phi$ of
post-fluidized states is history independent and Gaussian distributed about a stationary state. 
The standard deviation of $\phi$
exhibits, as a function of $\phi$, a minimum corresponding to a
maximum in the number of statistically independent regions.
Measurements of the fluctuations enable us to determine the
compactivity $X$, a temperature-like state variable introduced in the
statistical theory of Edwards and Oakeshott [Physica A {\bf 157}, 1080
(1989)].
\end{abstract}



\maketitle

Granular materials consist of a large number $N$ (typically more than
$10^6$) dissipative particles that are massive enough so that their potential
energy is orders of magnitude larger than their thermal energy.  The
large number suggests that a statistical description might be
feasible.  Edwards and coworkers \cite{edwards:89} developed such a
description with the volume $V$ of the system, rather than the energy,
as the key extensive quantity in a static granular system.
 The corresponding configuration space
contains all possible mechanically stable arrangements of grains.

Brownian motion is insufficient for a granular system to explore its
configuration space, so energy must be supplied by external
forcing such as tapping~\cite{nowak:98}, shearing~\cite{tsai:03}
or both~\cite{daniels:04}.
The theory of Edwards requires that the forcing assures ergodicity:
all mechanically stable configurations must be equally probable and
accessible.  A necessary condition for ergodicity is history
independence: the physical properties of the system must not depend on
the way a specific state was reached.  History independence has
previously been demonstrated only by Nowak {\it et
al.}~\cite{nowak:98} for tapped glass beads at volume fractions $\phi
> 0.625$.

In this paper we explore the configuration space using a periodic
train of flow pulses in a fluidized bed.  A stationary column of glass
beads in water is expanded by an upward stream of water until it
reaches a homogeneously fluidized state \cite{jackson:00}, and then
the flow is switched off.  The fluidized bed collapses
\cite{goldfarb:00} and forms a sediment of volume fraction $\phi$,
which we find depends in a reproducible way on the flow rate $Q$ of
the flow pulse.  This forcing results in a {\it history independent}
steady state where the volume exhibits Gaussian fluctuations around
its average value.

A central postulate of the Edwards theory is the existence of a
temperature-like state variable called compactivity $X = \partial V /
\partial S$.  The entropy $S$ is defined in analogy to classical
statistical mechanics as $S(V,N) = \lambda \ln{\Omega}$, where
$\Omega$ is the number of mechanically stable configurations of $N$
particles in $V$, and $\lambda$ is an unknown analog to the Boltzmann
constant.  The assumption that $X$ is a relevant control parameter in
granular systems has found support in simulations of segregation in
binary mixtures~\cite{nicodemi:02}, compaction under vertical
tapping~\cite{barrat:00}, and shearing~\cite{makse:02}.  However, no
measurements of $X$ have been reported.  In this paper we determine
$X$ from the measured volume fluctuations using a method suggested
in~\cite{nowak:98}.

\begin{figure}[b]
  \begin{center}
    \includegraphics[width=7.8cm]{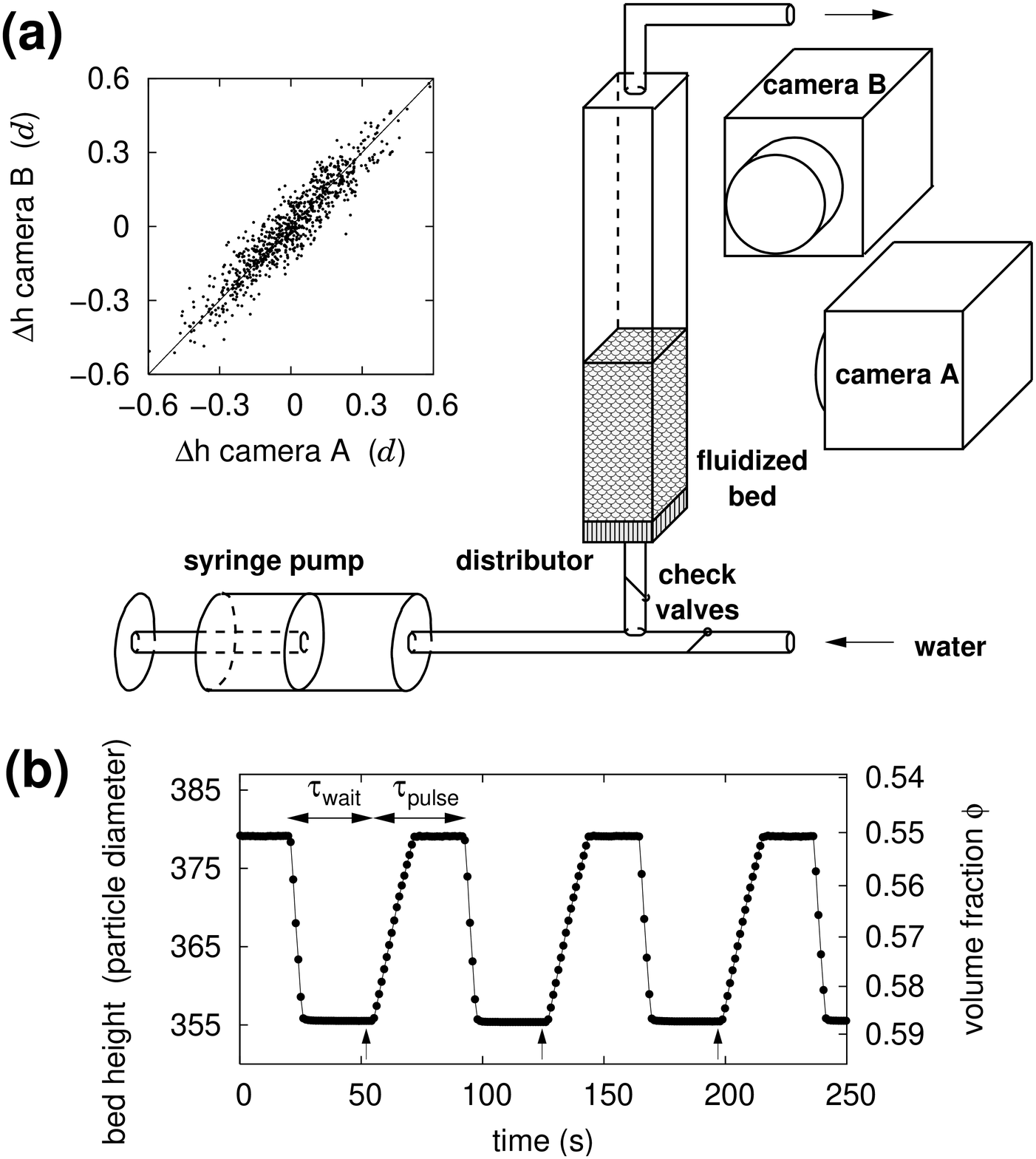}
    \caption{(a) Experimental setup.  The inset shows the correlation
      between the height changes $\Delta h$ during a flow pulse to the fluidized bed, 
	as seen by the two cameras (800 flow pulses, $Q$ = 60\,mL/min). (b) Development
      of the bed height during flow pulses of $Q$ = 40\,mL/min.  The
      arrows ($\uparrow$) indicate measurements of $h$.  }
    \label{fig:stack_puls}
  \end{center}
\end{figure}

{\it Experiment.} The apparatus is shown in Fig.~\ref{fig:stack_puls}(a).
A square bore glass tube (24.1\,mm $\times$ 24.1\,mm) contains about
$3.6 \times 10^6$ beads (soda-lime glass, $d=250 \pm 13~\mu$m, density
2.46\,g/cm$^3$, MO-SCI Corporation). The beads are fluidized with
pulses of temperature-controlled (23.0\,$\pm$\,0.1\,$^\circ$C)
de-ionized water. 

\begin{figure}[t]
  \begin{center}
    \includegraphics[angle=-90,width=8.6cm]{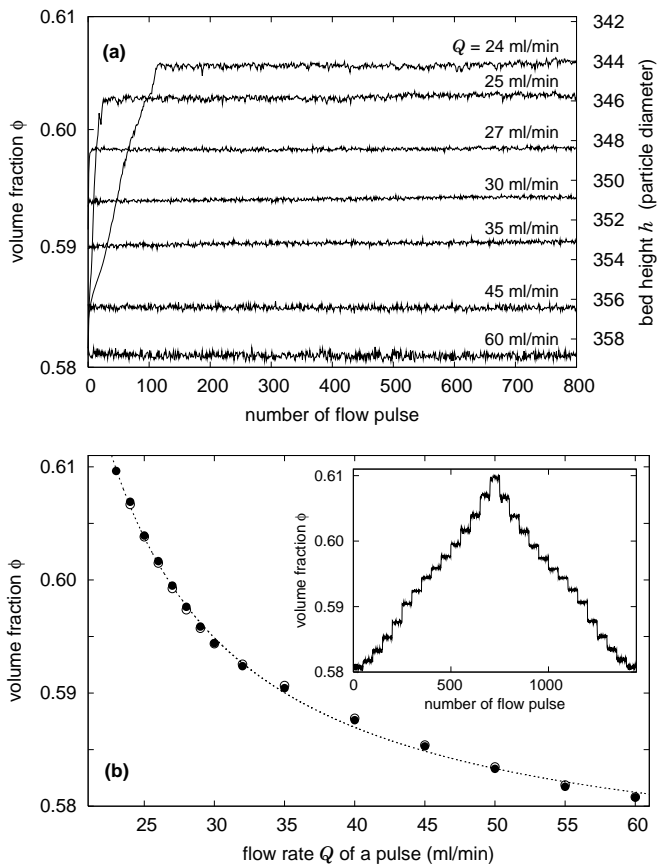}
    \caption{History independent steady state: (a) The volume fraction
        $\phi$ of the sedimented bed reaches a steady state that
        depends only on the flow rate $Q$. (b) The inset shows
        the increase in volume fraction as $Q$ was decreased in 14
        steps from 60\,mL/min to 23\,mL/min, followed by a decrease in
        volume fraction as $Q$ was increased again; each step
        consisted of 50 flow pulses.
	The main graph (b) shows that the steady state volume 
	fraction is the same
        for increasing ($\circ$) and decreasing ($\bullet$) $Q$; the
        dashed line is a fit to (\ref{eq:rlp}).
	}
    \label{fig:st_st_no_hi}
  \end{center}
\end{figure}

Flow pulses were generated by a computer-controlled syringe pump.
During a flow pulse of length $\tau_{\rm pulse}$
(Fig~\ref{fig:stack_puls}(b)), the bed expanded until its height
reached a stable value where the time-averaged viscous drag force on
each particle corresponded to its weight; $\tau_{\rm pulse}$ was
chosen so $Q \tau_{\rm pulse}$ = 30\,mL.
After each flow pulse, the bed was allowed to settle into a 
mechanically stable
configuration, which was determined by measuring the
time-resolved correlation of laser speckle patterns~\cite{cipelletti:03}.  A
waiting time $\tau_{\rm wait}$ of 30\,s was found to be ample
for achieving  a jammed state~\cite{liu:98}. After $\tau_{\rm wait}$
the volume fraction $\phi$ was determined by measuring the bed height $h$
with two CCD cameras at a 90$^\circ$ angle. Averaging over the width
of the bed yielded height values with a standard deviation of only
$0.008d$ for a bed at rest
(cf. inset of Fig.~\ref{fig:stack_puls}(a)).

{\it History independent steady state.} Starting with an initial $\phi$ =
0.581 prepared with a single flow pulse of $Q$ = 60\,mL/min, the volume
fraction quickly approaches a $Q$-dependent constant value $\phi_{\rm
  avg}$, as shown in Fig~\ref{fig:st_st_no_hi}\;(a). Then with
successive flow pulses, $\phi$ fluctuates about $\phi_{\rm avg}$.  The
history independence is demonstrated by ramping up and down in flow
rate (Fig.~\ref{fig:st_st_no_hi}(b)); $\phi$ depends only on $Q$ of
the last flow pulse, not the earlier history of the bed.  
The remaining differences in $\phi_{\rm avg}$ 
for increasing and decreasing $Q$ are on average only $1.6 \times 10^{-4}$.  
These small variations are
correlated with the viscosity changes due to the temperature drift of $\pm$0.1\,$^\circ$C 
during the course of the experiment (35\,h). 

For slow sedimentation, corresponding to $Q \rightarrow \infty$,
$\phi$ should converge to its random loose packing value, $\phi_{\rm
RLP}$ \cite{onoda:90}.  In the absence of a theory, we fit the data to
a phenomenological equation,
\begin{equation}
  \label{eq:rlp}
	\phi(Q) = \phi_{\rm RLP} + \frac{a}{Q - b}
\end{equation}	  
\noindent (Fig.~\ref{fig:st_st_no_hi}(b)).  This yields $\phi_{\rm
  RLP}$ = 0.573$\pm$0.001 ($a$= 0.365\,mL/min, $b$=12.9\,mL/min),
where the uncertainty arises not from the very small uncertainty in
bed height measurement (corresponding to 0.0003 in $\phi$) 
but from variations in the cross section of the square tube.

\begin{figure}[t]
  \begin{center}
    \includegraphics[angle=-90,width=8.6cm]{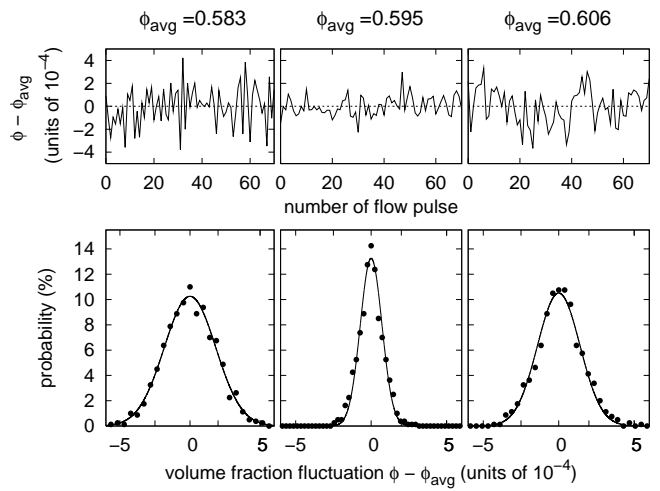}
    \caption{Volume fraction fluctuations around the steady states shown in 
      Fig.~\ref{fig:st_st_no_hi}(a) are Gaussian distributed.
      The upper row shows examples of the volume fraction measurements
      as a function of the number of flow pulses; $\phi_{\rm avg}$ is
      indicated above each column.  The lower row shows the
      corresponding probability distributions obtained from 800
      flow pulses; the curves are Gaussian fits.}
    \label{fig:multi_gauss}
  \end{center}
\end{figure}

{\it Volume fluctuations.} Gaussian fluctuations in $\phi$ are observed for
the whole range of $\phi$ studied; examples are shown in
Fig.~\ref{fig:multi_gauss}.  The standard deviation of these
fluctuations $\sigma_{\phi}$, like $\phi_{avg}$, was found to be
history independent.
\begin{figure}[t]
  \begin{center}
    \includegraphics[angle=-90,width=8.6cm]{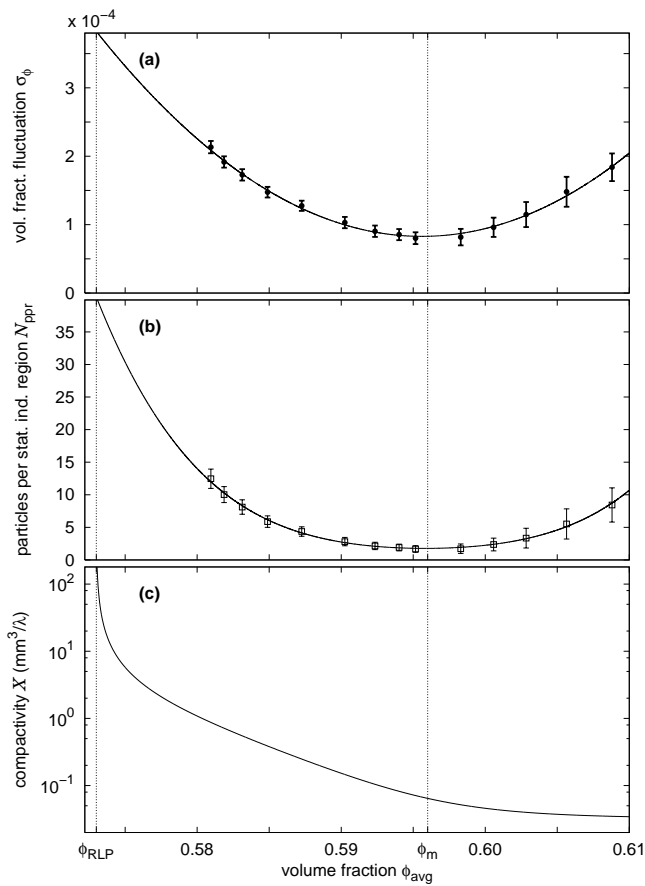}
    \caption{(a) Volume fraction fluctuation measurements fit a
        parabolic curve, (b) number of particles in a statistically
	independent region, 
	 and (c) compactivity as a function of the
        average volume fraction. At $\phi_{\rm m}$ the system has a maximum
        in the number of statistically independent regions
        (cf. (\ref{eq:stat_indep})). }
    \label{fig:sigma}
  \end{center}
\end{figure}
The variation of $\sigma_{\rm \phi}$ with $\phi$ fits a parabola with a
minimum at $\phi_{\rm m}$ = 0.596 (Fig.~\ref{fig:sigma}(a)).  The minimum
corresponds to a {\it maximum} in the number of statistically
independent spatial regions $N_{\rm i}$ at the moment of solidification,
\begin{equation}
  \label{eq:stat_indep}
       \frac{\sigma_{\phi}}{\phi_{\rm avg}} = \frac{k}{\sqrt{N_{\rm i}}},
\end{equation}	  
\noindent where $k \approx 0.2$,  estimated by Nowak et al.~\cite{nowak:98} 
from the maximal volume fraction fluctuations a single spatial region 
could undergo.
Using (\ref{eq:stat_indep}) and the parabolic
fit in Fig.~\ref{fig:sigma}(a), we estimate the average number of
particles per statistically independent region, $N_{\rm ppr}= N/N_{\rm
i}$ (where $N$ is the number of particles).
Figure \ref{fig:sigma}(b) shows $N_{\rm ppr}$ to be 40
for random loose packing, it reaches a minimum of about 1.8
at $\phi_{\rm m}$ and then increases again.  This contrasts with the roughly
constant value of $N_{\rm ppr}$ found in previous experiments~\cite{nowak:98} for
$\phi$ in the range 0.625-0.634.

We suggest that the minimum of $N_{\rm ppr}$ at $\phi_m$ corresponds
to a crossover of competing mechanisms.  With increasing $\phi$ above
$\phi_{RLP}$, where force chains stabilize the medium, the number of
contacts per grain increases \cite{silbert:02b}, and the probability
that the displacement of one grain will mechanically destabilize other
grains decreases. Consequently $N_{\rm ppr}$ decreases.  But with
increasing $\phi$, the free volume per particle decreases. 
The latter effect is dominant for $\phi > \phi_m$, 
where geometrical constraints allow movement only 
as  collective process of an increasing number of beads.
Hence $N_{\rm ppr}$ and
$\sigma_{\phi}$ increase.  At the maximum random jammed packing,
$\phi_{\rm MRJ}$ = 0.64, any motion would require a rearrangement of
the whole system~\cite{torquato:00}.

Knowing the dependency of $\sigma_{\rm \phi}$ on $\phi$ enables us to
determine Edward's compactivity $X$ using a granular version of the
fluctuation dissipation theorem derived by Nowak et
al.~\cite{nowak:98}. We obtain 
\begin{equation}
	\label{eq:gfdt}
	\frac{\lambda \rho}{m} \int_{\phi_{\rm RLP}}^{\phi} \left ( \frac{\phi'}{\sigma_{\phi'}}\right )^2  d\phi'=
	\frac{1}{X (\phi)}\quad,
\end{equation}
where we have used $X(\phi_{\rm RLP}) = \infty$ from granular
statistical mechanics, and $\rho$ is the particle density.  We solve
(\ref{eq:gfdt}) numerically by substituting $\sigma_{\phi}$ from the
parabolic fit (Fig.~\ref{fig:sigma}(a)) and starting at $\phi_{\rm
RLP}$ determined from the fit with (\ref{eq:rlp}).  This first
experimental result obtained for $X$ is shown in
Fig.~\ref{fig:sigma}(c); the values involve the unknown multiplicative
factor $\lambda$.  No theoretical or numerical work has predicted how
$X$ depends on $\phi$.  Knowing $X$ will allow further tests of
granular statistical mechanics.

\begin{figure}[t]
  \begin{center}
    \includegraphics[width=8.6cm]{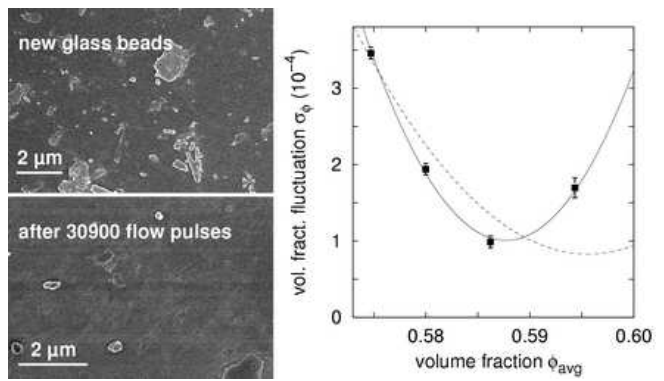}
    \caption{Influence of particle surface roughness.  On the left are scanning
      electron microscopy images of the surface of new and aged glass
      beads.  The plot on the right shows the volume fraction
      fluctuations of new beads ($\blacksquare$) together with a
      parabolic fit (solid line); to minimize aging, measurements were
      made with only 200 flow pulses.  The dashed line is the
      parabolic fit for aged beads from Fig.~\ref{fig:sigma}.}

    \label{fig:surface}
  \end{center}
\end{figure}

{\it Aging and friction.}
Our experiments revealed an aging effect due to a change in the
frictional properties of the beads.  Friction between beads decreases
with multiple collisions, which lead to a decrease in the number of
surface asperities, as illustrated in Fig.~\ref{fig:surface}. 
The data we have presented show no aging during the 12,000 pulses in the experiment;
the beads had been previously used for more than 
45,000 flow pulses. Just as for aged beads, the volume
fraction standard deviation for new beads fits a parabola, as
Fig.~\ref{fig:surface} illustrates; the minimum is at $\phi_{\rm m}$=0.587.
The shift of the minimum to a value lower than lower $\phi_{\rm m}$=0.596 for aged beads 
is consistent with our argument
about statistically independent regions: a higher values of $\mu$ makes
dilute states less fragile and lowers the onset of the necessity of
collective motion. For new beads a fit to (\ref{eq:rlp}) results in
$\phi_{\rm RLP} = 0.568 \pm 0.001$, in accord with $\phi_{\rm RLP} =
0.566 \pm 0.004$ interpolated from the measurements (presumably with new beads)
in~\cite{onoda:90}. 
This shows that $\phi_{\rm RLP}$ depends on the frictional properties 
of the beads, as predicted in \cite{srebro:03}.

{\it Conclusions.} We have shown that the configuration space of a granular
medium can, by using flow pulses, be explored in a history independent
way, which is essential for the statistical theory~\cite{edwards:89}
to be applicable.  Fluctuations of the volume fraction are Gaussian
with a parabolic minimum, which corresponds to a maximum in the number
of statistically independent regions. The minimum arises as a
consequence of competing mechanisms; the location of the minimum
depends on the frictional properties of the beads.  The compactivity
can be determined from the volume fraction fluctuations; hence
compactivity is a well-defined parameter representative of the
material.  This opens the door for new experiments that would, for
example, investigate the use of compactivity as a control parameter
for segregation in binary mixtures~\citep{nicodemi:02,srebro:03}.

We thank Antonio Coniglio for rousing our interest in this problem;
Massimo Pica Ciamarra, Karen Daniels, Sunghwan Jung, Salvatore
Torquato, and Thomas Truskett for helpful discussions; and Michael
Schmerling for help with the SEM. This work was supported by the
Engineering Research Program of the Office of Basic Energy Sciences of
the U.S. Department of Energy (Grant No. DE-FG03-93ER14312), by the
Robert A Welch Foundation, and by the Office of Naval Research Quantum
Optics Initiative (Grant No. N00014-03-1-0639).


\end{document}